\documentclass[12pt]{article}
\usepackage{mathtools}
\newcommand{\fsl}[1]{\ensuremath{\mathrlap{\!\not{\phantom{#1}}}#1}}

\begin{document}

\begin{center}
  {\large \bf On  Semiclassical  Equivalence of G-S and Pure Spinor Strings  in $AdS(5)\times S(5)$ }

\vspace{5mm}

\end{center}
\begin{center}
Mario Tonin \footnote{e--mail: tonin@pd.infn.it}

\vspace{0.5cm}
{\it Universit\`a Degli Studi Di Padova,
Dipartimento Di Fisica e Astronomia ``Galileo Galilei''\\
ed  \\
INFN, Sezione Di Padova,
Via F. Marzolo, 8, 35131 Padova, Italia}  \\
\end{center}
\vspace{1.cm}
\begin{center}
{\bf Abstract}
\end{center}
We present a method to study  the equivalence at the semiclassical level of the Green-Schwarz and pure spinor
formulations of String Theory in $AdS(5)\times S(5)$. This method provides a clear separation of the physical and unphysical sectors of the pure spinor formulation and allows to prove that the two models  have not only equal spectra of the fermionic fluctuations but  also equal conformal weights (the bosonic ones being equal  by construction).

\thispagestyle{empty}
\newpage
\section{Introduction}
String theory  in $AdS(5)\times S(5)$ is of great  relevance being the first  and  well  studied example of
AdS/CFT correspondence. The Ramond--Neveu--Schwarz formulation  is not suitable  to  describe  string theory  in this  background due  to  the  presence  of a non--vanishing  Ramond--Ramond flux. Two approaches  are available in this case:  the  Green-Schwarz (GS) formulation \cite{gs} and  the  pure spinor (PS) one \cite{berk}.
The GS formulation for Type II superstrings in a general curved background is known for a long time \cite{Grisaru:1985fv}. The PS formulation in a generic superbackground has been  constructed in \cite{berk-howe} by  considering  the  more general  conformal  invariant action involving  also the  momenta $ d_{L,\alpha}$, $ d_{R, \alpha}$ of the Grassmann--odd superspace  coordinates $\theta_{L/R}^{\alpha}$ and  the  ghosts $\lambda_{L/R}^{\alpha}$,
$ \omega_{L/R,\alpha}$. The  BRS  invariance of the PS action  implies the  on shell  supergravity  constraints and  holomorphicity  of BRS currents. For further discussions  see \cite{gut},\cite{ch},\cite{bed}. Alternatively, starting from an extended free differential algebra of the superspace geometry \cite{dauria},
\cite{ton}, \cite{ot2},
one  can add to  the GS action  conformal  invariant  terms  involving  $d_{L/R}$  and  ghosts  to  promote  the $\kappa$--symmetry of  the GS approach  to  BRS  symmetry \cite{ton},\cite{ot2} following a method proposed in \cite{ot1} for the heterotic case.

The GS  and  PS actions in  $AdS(5)\times S(5)$ are  obtained  from  the corresponding actions in a generic curved superbackground by  setting  the background  superfields  to  their  $AdS(5)\times S(5)$ values.
An incomplete list of papers that discuss the  $AdS(5)\times S(5)$  string theory are  \cite{mets}-\cite{tseyt} in the GS formulation and \cite{berkch}--\cite{maz} in the pure spinor formulation.
In  the  PS case  there  is a term, proportional  to  $(d_{L} M d_{R})$  where  M is a superfield  related  to  the  R-R  flux. In
$AdS(5)\times S(5)$ M is a constant invertible  matrix and, since  the  other  terms  are  almost  linear  in  $d_{L/R}$, one can  integrate  over $d_{L}$  and  $d_{R}$ to  get  an action that  depends  on  the  same
 superspace  variables  as in the GS  model  (in  addition  to the ghosts). Since $AdS(5)\times S(5)$ is  the  coset  $ {PSU(2,2/4) \over {SO(4,1)\times SO(5)}} $  it  is convenient  to  express  the GS and PS  actions  in  terms  of  the  currents, valued  in  this  coset \cite{mets}. The bosonic  sectors  of  the GS and  PS  actions  are equal but  the  fermionic sectors  are  different. In particular, in the PS formulation the fermionic modes have a second--order kinetic term. Then  a natural  question to  ask  is whether  the  two  formulations  are equivalent.

It  does not seem easy to answer  this  question  in general.
In a  flat  background, the quantum  equivalence  of  the  GS superstring  (in semi--light--cone  gauge) and  the  PS  superstring has  been proven  in \cite{bm,ak} long time  ago. That  has  been  possible  essentially  because in  flat  background  string  theories  are free  field  theories on  the  world--sheet.  There have  been  attempts  to  relate these  two  approaches also  in a curved  background, but  only  at  the  classical  level. For  instance, in \cite{mmost}, the classical  PS  formulation (for  the  heterotic string)  was  obtained  from a  twistor--like (or  superembedding) formulation  which is  equivalent  to  the  GS  one.  To  prove this  equivalence at  the  full  quantum  level  in  a  curved  background  (where  the  strings  are  not  free)  seems  an  almost  hopeless  problem.

However  it  is  possible to study  this  problem   in a simplified  set up, that is  at the semiclassical level: one expands the GS and PS actions  around a classical solution  of  the  bosonic field equations  of the  string up  to terms quadratic in  the  fluctuations  and  then compares  their (fermionic)  spectra (equivalently  their one-loop (fermionic)  partition functions). In $AdS(5)\times S(5)$  background, this  problem has been faced  in two recent  papers \cite{abv,dima}. In \cite{abv}
the equality  of the fermionic  spectra  has  been  proven for  a  simple family  of  string  motions  and  in \cite{dima} this  result  has  been  extended to a generic  motion  of  the  string  in $AdS(5)\times S(5)$.

In this  paper  we  propose  a different  method  to  deal  with this  problem at  the semiclassical level.
The  method  has  the  advantage  that it  provides a clear  separation of  the  physical  and unphysical sectors of the PS approach  and allows to prove not  only  the equivalence of  the  spectra  of the physical fermionic  fluctuations of the GS and PS formulations but  also  their conformal  weights. Indeed  it  is  shown  that  the  PS model  in the considered approximation contains $8_L + 8_R$ fermions with  conformal  weight (0,0) and  the  same ``mass"  (in general world--sheet dependent) as in the GS approach, and  $2_L + 2_R$ massless fermions  with  conformal  weights
(2,0), (-1,0), (0,2), (0,-1) that  match the  masses and the conformal weights of the b-c ghosts, which are present in the GS approach to  fix  the  world--sheet  diffeomorphisms in the conformal  gauge.

The  unphysical (bosonic and fermionic) fluctuations  give  rise  to eleven  left--handed  and  eleven  right--handed massless BRS quartets so  that  their  contributions to  the  one--loop  partition  function  cancel.

\section{GS Action in $AdS(5)\times S(5)$ }

In the conformal gauge, in a generic curved background the GS action  is
\begin{eqnarray}
I_{GS} =  {{R^2} \over {\pi \alpha'}}\int dz^{+}dz^{-}[ E_{+}^{a}(Z)E_{- a}(Z) + B_{+ -}(Z)]
\label{01}
\end{eqnarray}
Where $ E^{A}(Z) \equiv (E^{a}, E_{L}^{\alpha}, E_{R}^{\alpha})$ are the vector-like ($ E^{a}$, $ a = 0,...,9$) and spinor-like ($ E_{L/R}^{\alpha}$, $\alpha=1,\ldots, 16$) supervielbeins and  B is the NS-NS two-superform. $E_{\pm}^{a}$  and $B_{+ -}$ are the pull-backs of these  forms onto the world sheet parametrized by the coordinates $z^{\pm}$, and $Z^{M} \equiv (x^{m}, \theta_{L}^{\mu}, \theta_{R}^{\mu})$ are the target superspace  coordinates. If the  superspace geometry  satisfies  the (on-shell) supergravity  constraints, the action (\ref{01}) is invariant  under $\kappa$--symmetry
$$
\delta Z^{M} E_{M, L}^{\alpha} =  (E_{-}^{a}\Gamma_{a} k_{L})^{\alpha} \qquad  \delta Z^{M} E_{M, R}^{\alpha} =  (E_{+}^{a}\Gamma_{a} k_{R})^{\alpha}
$$
($k_{L/R}$ being  local parameters) that  halves the number  of fermionic degrees of freedom.
In $AdS(5)\times S(5)$  $E^{a}$ decomposes as $ E^{a} \equiv ( E^{\tilde a},E^{i})$ where $\tilde a = 0,1,..4$ refers to $AdS(5)$  and $i = 5,...9 $ refers to $S(5)$. A similar decomposition  holds  for
the Lorentz  connection $ \omega^{[ab]} \equiv( \omega^{[\tilde a \tilde b]}, \omega^{[ij]})$ with  curvature
\begin{eqnarray}
 R^{abcd} = (- R \eta^{\tilde a [\tilde b}\eta^{\tilde c] \tilde d}, \, R \delta^{i [j}\delta^{h],k}))
\label{01bis}
\end{eqnarray}
We will assume  $R=1$ in the  following. Moreover in this background the only  non--vanishing  component
of  B is \cite{berkbis} $ (E_{L}B^{(o)}E_{R})$ where, with R=1,
\begin{eqnarray}
 B^{(o)}_{\alpha \beta} = (\gamma^{01234})_{\alpha\beta} \equiv (\gamma_{\ast})_{\alpha\beta}
\label{02}
\end{eqnarray}
and $$(\gamma_{\ast})^2 = - 1 .$$ The matrix $\gamma_{\ast}$ is also equal to the constant matrix $M^{(o)}$ that describes the R-R flux in this background:
\begin{eqnarray}
M^{(o)} = \frac{1}{2\cdot 5!} F^{abcdf}\gamma_{abcdf}= \gamma_{\ast}
\label{03}
\end{eqnarray}
Then the GS action in $AdS(5)\times S(5)$ is
\begin{eqnarray}
I_{GS} = \int L_{GS} = \int [ E_{+}^{a}(Z)E_{- a}(Z) + {1 \over 2}( E_{L,+}(Z)\gamma_{\ast}E_{R,-}(Z)) - {1 \over 2}(E_{L,-}(Z)\gamma_{\ast}E_{R,+}(Z))]
\label{04}
\end{eqnarray}
Here and in the following  $\int..$  stands for $\int {{dz^{+}dz^{-}}\over {\pi \alpha'}}..$.

Since $AdS(5)\times S(5)$ is the coset $$ {G \over H} = {{PSU(2,2/4)} \over { SO(4,1)\times SO(5)}} $$
it is convenient  to  express this action  in  terms of the  currents
\begin{eqnarray}
g^{-1}d g = J \equiv J^{0} + J^{1} + J^{2} + J^{3}
\label{05}
\end{eqnarray}
where g is an element of $ G  $ that transforms under $ G$ from the left and under the structure group $H$ on the right and $ J = J^{A} T_{A}$ is valued  in the Lie algebra of $ G $,  $ T_{A}$ being the generators of  $psu(2,2/4)$, that is  $ J^{2} = J^{2,a}T_{a}$, $ J^{1} = J^{1, \alpha}T_{L, \alpha}$ $ J^{3} = J^{3,\alpha}T_{R,\alpha}$,  $ J^{0} = J^{0,[ab]}T_{[ab]}$ . As shown in \cite{berkbis}, in terms of these currents  the action (\ref{04}) can be written as
\begin{eqnarray}
 I_{GS} = {1 \over 2} \int  str[ J_{+}^{2}J_{-}^{2} - {1 \over 2 }( J_{+}^{1}J_{-}^{3} -
 J_{+}^{3}J_{-}^{1})]
 \label{06}
\end{eqnarray}
where $ str $ denotes  the  supertrace.
This form  of the action  is  useful since  it  reveals the hidden $Z_{4}$ automorphysm of the $ AdS(5)\times S(5)$ action, the index r in $J^{r}$ being its grading under $Z_{4}$. The relation with the  superspace notations of (\ref{01})--(\ref{04}) is:
$ J^{2,a} = E^{a}$, $J^{1, \alpha}= E_{L}^{\alpha}$ , $J^{3, \alpha}= E_{R}^{\alpha}$ and $ J^{0,[ab]} =
\omega^{[ab]} $ .

We are interested in studying the  motion of a string in $AdS(5)\times S(5)$ at the semiclassical level, i.e. to compute the spectra  of the  quantum  fluctuations around the classical solution. For that one considers the $AdS(5)\times S(5)$ GS action  expanded around a classical  solution  of the bosonic field  equations. A generic classical motion of the bosonic string is described by  the pull-back of the classical vielbeins $ e_{+}^{a}$, $e_{-}^{a}$ which in the conformal gauge satisfy  the  field equations
\begin{eqnarray}
 \nabla_{+}e_{-}^{a} = 0  = \nabla_{-}e_{+}^{a}
 \label{07}
 \end{eqnarray}
 and  the Virasoro constraints
 \begin{eqnarray}
 e_{-}^{a}e_{- a} = 0  =  e_{+}^{a}e_{+ a}
 \label{08}
 \end{eqnarray}
$\nabla_{\pm} $ are the pull-back of the  covariant derivative $ \nabla = d + \omega $ where
$ \omega^{[ab]}$ is the classical  Lorentz connection considered previously. It  will be convenient  to define
\begin{eqnarray}
 \fsl{e_{\pm}} =
e_{\pm}^{a}\Gamma_{a}\equiv  e_{\pm}^{\bar a}\Gamma_{\bar a} + e_{\pm}^{i}\Gamma_{i}
\label{09}
\end{eqnarray}
Expanding eq.(\ref{04}) (or (\ref{06})) around this  classical  solution up to terms quadratic in fluctuations, the action  decomposes into a bosonic part and a fermionic part. Since, as already noted and as will be seen below, the bosonic parts are the same for the GS and PS formulations, for the sake of comparison of the two approaches, one can forget the bosonic parts. The fermionic part of the GS Lagrangian, expanded up to  quadratic terms, is
\begin{eqnarray}
L_{GS} ={1 \over 2} (\theta_{L}\fsl{e_{-}}\nabla_{+}\theta_{L})   + {1 \over 2} (\theta_{R} \fsl{e_{+}}\nabla_{-}\theta_{R}) - {1 \over 2}( \theta_{L}\fsl{e_{-}}\gamma_{\ast} \fsl{e_{+}}\theta_{R})
\label{010}
\end{eqnarray}
which is  invariant under  the simplified $\kappa$--symmetry
$$ \delta \theta_{L/R} = (\fsl{e_{\mp}}k_{L/R}). $$ If one defines the fermions
$$  \theta^{(\pm)}_{L/R} =  P_{\pm}\theta_{L/R}, $$ where $ P_{\pm}$ are the projectors
\begin{eqnarray}
P_{\pm} = {1 \over 2(e_{+} e_{-})}\fsl{e_{\pm}} \fsl{e_{\mp}}  \qquad  P_{+} + P_{-} = 1
\label{30}
\end{eqnarray}
that  project on an 8--dimensional  subspace of the 16 dimensional  spinorial  space,  the  fermions $  \theta^{(\pm)}_{L/R} $ have 8 components each. The $\kappa$--symmetry implies  that (\ref{8})  does  not  depends  on  $ \theta^{(-)}_{L}$  and $ \theta^{(+)}_{R}$
so  that  the  Lagrangian
\begin{eqnarray}
L_{GS} ={1 \over 2} (\theta^{(+)}_{L}\fsl{e_{-}}\nabla_{+}\theta^{(+)}_{L})   + {1 \over 2} (\theta^{(-)}_{R} \fsl{e_{+}}\nabla_{-}\theta^{(-)}_{R}) - {1 \over 2}( \theta^{(+)}_{L}\fsl{e_{-}}\gamma_{\ast} \fsl{e_{+}}\theta^{(-)}_{R})
  \label{011}
  \end{eqnarray}
describes  8 left-handed  and 8  right-handed massive  fermions with conformal weights (0,0), which are the  fermionic partners of  the 8 transverse bosons $x^{\hat m}$. The values of the fermion masses are determined by the last term of \eqref{011}. Notice that since  this term  involves  the  classical vielbeins  $e_{\pm}(x(z))$, in general the ``masses" depend on the world--sheet coordinates $z^{\pm}$.

Let us also recall that fixing the world--sheet diffeomorphisms by imposing the conformal gauge requires the GS action to be supplemented with the b-c ghost action
\begin{eqnarray}
\int L_{ghost} =  \int [ b_{L,--}\partial_{+}c^{-}_{L} + b_{R,++}\partial_{+}c^{+}_{R}]
\label{012}
\end{eqnarray}
where  the  ghosts $b_{L,--}$, $c_{L}^{+}$ and $b_{R,++}$, $c_{R}^{-}$  have  conformal  weights (2,0), (-1,0) and (0,2), (0,-1) respectively.

\vspace{5mm}
\section{\bf PS Action in $AdS(5)\times S(5)$ }
\vspace{2mm}

As said before, the  pure spinor action in a generic curved background is obtained by adding to the GS action conformal invariant terms involving the  fermionic  momenta $ d_{L/R}$ and the pure--spinor ghosts, so that the PS action becomes invariant under BRS transformations generated by the BRS charge
\begin{eqnarray}
 Q_{BRS} = \int dz^{-}(d_{-,L}\lambda_{L}) + \int dz^{+}(d_{+,R}\lambda_{R})
 \label{12bis}
 \end{eqnarray}
if the supergravity background is on shell. In $AdS(5)\times S(5)$
the PS action takes the form
\begin{eqnarray}
I = \int  L_{GS} + \int  [ (E_{L,+}d_{L,-}) + (E_{R,-}d_{R,+})  - (d_{L,-}\gamma_{\ast}d_{R,+})]
\nonumber
\end{eqnarray}
\begin{eqnarray}
+ I_{\omega\lambda}
\label{013}
\end{eqnarray}
where
\begin{eqnarray}
I_{\omega\lambda} = \int [ (\omega_{L, -}\nabla_{+} \lambda_{L}) + (\omega_{R, +}\nabla_{-} \lambda_{R}) + R_{abcd}(\omega_{L,-}\Gamma^{a[b}\lambda_{L}) (\omega_{R,+}\Gamma^{c]d}\lambda_{R})]
\label{014}
\end{eqnarray}
with $L_{GS}$  defined in (\ref{04}) and $ R_{abcd} $ in (\ref{01bis}). $ \lambda_{L/R}$ are pure spinors satisfying the constraints
$$ (\lambda_{L}\Gamma^{a}\lambda_{L}) = 0 = (\lambda_{R}\Gamma^{a}\lambda_{R})$$
 and $I_{\omega\lambda}$
is invariant under the $\omega$-symmetry
$$ \delta \omega_{L/R} = \Lambda_{L/R}^{a}\Gamma_{a}\lambda_{L/R}$$ where $\Lambda_{L/R}^{a}$ are local parameters.
Eq. (\ref{013}) can be integrated over $d_{L/R}$ and expressed  in  terms  of the currents (\ref{05}), as in the
GS case. The  result is \cite{berkch}
\begin{eqnarray}
 I_{PS} = {1 \over 2} \int  str[ J_{+}^{2}J_{-}^{2} +{1 \over 2} J_{+}^{1}J_{-}^{3} + {3 \over 2}
 J_{+}^{3}J_{-}^{1}]  + I_{\omega\lambda}
 \label{015}
\end{eqnarray}
As anticipated and as it is now clear from (\ref{06}) and (\ref{015}), the physical bosonic sectors in the GS and PS approaches are identical. Therefore, in order to compare the two approaches at the semiclassical level, it is sufficient to consider the fermionic sector of (\ref{013}) or (\ref{015}), expanded around the bosonic solution  considered in (\ref{07}), (\ref{08}), up to terms quadratic in the fluctuations.
For the purposes of the present paper it is convenient to start from the non integrated action (\ref{013}).
In this approximation the Lagrangian for the fermionic  sector  is
\begin{eqnarray}
L = {1 \over 2} [ \theta_{L}\fsl{e_{-}}(\nabla_{+}\theta_{L} - {1 \over 2} \gamma_{\ast} \fsl{e_{+}}\theta_{R}) + \theta_{R} \fsl{e_{+}}(\nabla_{-}\theta_{R} + {1 \over 2} \gamma_{\ast} \fsl{e_{-}}\theta_{L})]
 \nonumber
 \end{eqnarray}
 \begin{eqnarray}
 - [ d_{L}(\nabla_{+}\theta_{L} - {1 \over 2} \gamma_{\ast} \fsl{e_{+}}\theta_{R}) + d_{R}(\nabla_{-}\theta_{R} + {1 \over 2} \gamma_{\ast} \fsl{e_{-}}\theta_{L}) + {1 \over 2} (d_{L}\gamma_{\ast}d_{R}) ]
 \nonumber
 \end{eqnarray}
 \begin{eqnarray}
  + [(\omega_{L, -}\nabla_{+} \lambda_{L}) + (\omega_{R, +}\nabla_{-} \lambda_{R})]
\label{1}
\end{eqnarray}
Upon integrating ({\ref{1}}) over $d_{L}$ and $d_{R}$ one gets
\begin{eqnarray}
L = {1 \over 2} [ (\theta_{L} \fsl{e_{-}}(\nabla_{+}\theta_{L} - {1 \over 2} \gamma_{\ast}  \fsl{e_{+}}\theta_{R})) + (\theta_{R} \fsl{e_{+}}(\nabla_{-}\theta_{R} + {1 \over 2} \gamma_{\ast} \fsl{e_{-}}\theta_{L}))]
 \nonumber
 \end{eqnarray}
 \begin{eqnarray}
 - 2[(\nabla_{+}\theta_{L} - {1 \over 2}\theta_{R}\fsl{e_{+}}\gamma_{\ast})\gamma_{\ast} (\nabla_{-}\theta_{R} + {1 \over 2} \gamma_{\ast} \fsl{e_{-}}\theta_{L}))]
 + [(\omega_{L, -}\nabla_{+} \lambda_{L}) + (\omega_{R, +}\nabla_{-} \lambda_{R})]  =
 \nonumber
 \end{eqnarray}
 \begin{eqnarray}
 = - 2 (\nabla_{+}\theta_{L}\gamma_{\ast}\nabla_{-}\theta_{R}) - {1 \over 2}(\theta_{L} \fsl{e_{-}}\nabla_{+}\theta_{L})  - {1 \over 2}(\theta_{R}\fsl{e_{-}}\nabla_{-}\theta_{R})
\nonumber
\end{eqnarray}
\begin{eqnarray}
+ [(\omega_{L, -}\nabla_{+} \lambda_{L}) + (\omega_{R, +}\nabla_{-} \lambda_{R})]
\label{1bis}
\end{eqnarray}
which  is  the quadratic string  action  considered in \cite{abv,dima}.

Let us come back to eq.({\ref{1}}). Let us define the projectors   \footnote { In  $AdS(5)\times S(5)$,  $ (\lambda_{R}\gamma_{\ast}\lambda_{L})  $ belong to the BRS cohomology and  can be assumed to be non vanishing \cite{berkquater}.  }

\begin{eqnarray}
K_{L} = {1 \over 2} (\Gamma^{a}\gamma_{\ast}\lambda_{R}){1 \over {(\lambda_{R}\gamma_{\ast}\lambda_{L})}}
(\lambda_{L}\Gamma_{a})
\label{2}
\end{eqnarray}
\begin{eqnarray}
K_{R} = {1 \over 2} (\Gamma^{a}\gamma_{\ast}\lambda_{L}){1 \over {(\lambda_{R}\gamma_{\ast}\lambda_{L})}}
(\lambda_{R}\Gamma_{a})
\label{3}
\end{eqnarray}
and $ \tilde K_{L}$ and $ \tilde K_{R}$ are the  transposed of $ K_{L}$ and $K_{R}$.
Since $ TrK_{L/R} = 5 $, $ K_{L/R}$ and $ 1 - K_{L/R}$  decompose the 16-dimensional spinorial space
into  5-dimensional and  11-dimensional  subspaces, respectively.
Notice that
\begin{eqnarray}
 K_{L} \lambda_{L}= 0 = K_{R}\lambda_{R}
 \label{3a}
\end{eqnarray}
  so that  the  pure  spinors $ \lambda_{L/R}$  have  11 components. Moreover
\begin{eqnarray}
\lambda_{L}\Gamma^{a}(1 - K_{L}) = 0 =  \lambda_{R}\Gamma^{a}(1 - K_{R})
 \label {3b}
 \end{eqnarray}
 Other useful identities are
 \begin{eqnarray}
 \tilde K_{L} + \gamma_{\ast}K_{R}\gamma_{\ast}= 0 = \tilde K_{R}+ \gamma_{\ast}K_{L}\gamma_{\ast}
 \label{3c}
 \end{eqnarray}
 \begin{eqnarray}
 (\tilde K_{L}\Gamma^{a}K_{L}) = 0 =  ( \tilde K_{R}\Gamma^{a}K_{R})
 \label{3e}
 \end{eqnarray}
 and, from $(K_{L})^2 = K_{L}$,
 \begin{eqnarray}
 K_{L} (\nabla K_{L})K_{L} = 0 = (1 - K_{L}) (\nabla K_{L})(1 - K_{L})
 \label{3f}
 \end{eqnarray}
 and the same  for  $K_{R}$.
 Moreover, by gauge  fixing the $\omega$--symmetry one can impose the conditions
  \begin{eqnarray}
  \omega_{L}K_{L}= 0 = \omega_{R}K_{R}
  \label{3g}
  \end{eqnarray}

In order to avoid problems with the semiclassical approximation it is convenient  to  assume  that $\lambda_{L/R}$  decompose as $ \lambda_{L/R} = \lambda_{0,L/R} + \hat \lambda_{L/R} $ where $ \lambda_{0, L/R} $ are classical  fields subjects to the field equations $ \nabla_{\mp}\lambda_{0,L/R} = 0 $ and $ \lambda_{0,L/R} $, $\hat \lambda_{L/R} $ are pure spinors. Then  $K_{L/R}$ become $$ K_{L} = {1 \over 2} (\Gamma^{a}\gamma_{\ast}\lambda_{0, R}){1 \over {(\lambda_{0, R}\gamma_{\ast}\lambda_{0,L})}}
(\lambda_{0, L}\Gamma_{a}) + O(\hat \lambda). $$  Notice  that now in (\ref{1}) ( and in (\ref{1bis})) one must  add the term $$ - (\omega_{L,-}[R_{abcd} \Gamma^{a[b}\lambda_{0,L}) (\lambda_{0,R}\Gamma^{c]d}]\omega_{R, +}) $$ coming  from  the  last  term of (\ref{014}). This  term does not  affects the masslessness of the ghosts $ \lambda_{L/R}$ and $\omega_{L/R} $.

The action based  on the Lagrangian ({\ref{1}}) is  invariant under the BRS transformations
\begin{eqnarray}
s  \nabla_{+}\theta_{L} = \nabla_{+} \lambda_{L} \qquad
s \nabla_{-} \theta_{R} = \nabla_{-}\lambda_{R}
\label{4}
\end{eqnarray}
\begin{eqnarray}
s  \theta_{L}\fsl{e_{-}} = \lambda_{L} \fsl{e_{-}}  \qquad
s \theta_{R}\fsl{e_{+}} = \lambda_{R}\fsl{e_{+}}
\label{4bis}
\end{eqnarray}
\begin{eqnarray}
s \omega_{L} = - (d_{l,-}(1 - K_{L})) \qquad
s \omega_{R} = - (d_{R,+}(1 - K_{R}))
\label{5}
\end{eqnarray}
\begin{eqnarray}
s d_{L, -} = (\lambda_{L}\fsl{e_{-})} \equiv (\lambda_{L}\fsl{e_{-}})K_{L}  \qquad
s d_{R, +} = (\lambda_{R}\fsl{e_{+}}) \equiv (\lambda_{R}\fsl{e_{+})}K_{R}
\label{6}
\end{eqnarray}

Notice  that  we  abstain  ourselves  to  define the BRS transformations  of  $\theta_{L/R}$  themselves, i.e. $ s \theta_{L/R} = \lambda_{L/R} $, since  in a curved  background as $ AdS(5)\times S(5) $,
$ \lambda_{L/R}^{\alpha} $   transform  as  spinors  under the  target--space structure  group and $\theta_ {L/R}^{\mu}$ transform as odd  superspace  coordinates. Only  the  space--time superfields  or  forms  like  $E^{A}_{L/R} $, B,  etc.  have definite  transformations  properties  under  the BRS symmetry.   In  addition  notice  that, allowing  the BRS  transformations   $s \theta = \lambda$, we would obtain  that $ s {{(\theta_{L}\gamma_{\ast}\lambda_{R})} \over {(\lambda_{L}\gamma_{\ast}\lambda_{R})} }  =  1 = s {{(\theta_{R}\gamma_{\ast}\lambda_{L})} \over {(\lambda_{L}\gamma_{\ast}\lambda_{R})} } $, which    trivializes  the  cohomology.
Then  we  never  use  them  to  study  the  cohomology  of  our  model. But since  in  our  approximation  the  model  is  free, for all  other  instances the  use  of
$s \theta_{L/R} = \lambda_{L/R}$ is  safe.

Then from ({\ref{4}}), (\ref{4bis}) and ({\ref{5}}) it  follows that $11_L+11_R$ components $ (1 - K_{L})\theta_{L}$ and $ (1 - K_{R})\theta_{R}$, as well as
$\omega_{L}$ and $\omega_{R}$, are  not BRS invariant and  therefore  the  states involving these  fields
are not present in  the physical Fock space which is  defined as $$ |\psi> \in {\cal F}_{ph} \subset {\cal F}\qquad iff  \qquad Q_{BRS}|\psi> = 0$$ where ${\cal F}$  is the Fock space  of  the system. The physical  space  is  defined as
$$ {\cal H}_{ph} =  { {Ker(Q_{BRS})_{{\cal F}}} \over {Im(Q_{BRS})_{{\cal F}}}}. $$
At the same time
  \begin{eqnarray}
  \theta^{ph}_{L} = K_{L}\theta_{L}, \qquad  \theta^{ph}_{R} = K_{R}\theta_{R}
   \label{7}
   \end{eqnarray}
   are BRS invariant and  are physical fields. Notice that, since $e_{\pm}^{a}$ are
classical  fields, it follows from ({\ref{6}}) that $ d_{L,-}K_{L}$ and $ d_{R,+}K_{R}$ are not invariant  under BRS.  However if one  defines  the  combinations
\begin{eqnarray}
\hat d_{L, -} = d_{L,-} - \theta_{L}\fsl{e_{-}},  \qquad  \hat d_{L, -} = d_{R,+} - \theta_{R}\fsl{e_{+}}
\label{7a}
\end{eqnarray}
then
\begin{eqnarray}
d_{L,-}^{ph}=\hat d_{L, -}K_{L},\qquad   d_{R,+}^{ph}=\hat d_{R, +}K_{R}
 \label{7b}
 \end{eqnarray}
 are physical, being BRS invariant but not BRS exact. They can be  considered as the conjugate momenta of  $ \theta^{ph}_{L}$ and $ \theta^{ph}_{R}$.

Here a comment  is  in order.  The appropriate frame to work in the pure spinor approach is the Wick rotated Euclidean version of the theory  with $SO(10)$ as the structure group. The pure spinors (as well as any spinor projected with $K_{L}$, $K_{R}$, $(1 - K_{L})$, $(1 - K_{R})$ and  any  tensor  of $SO(10)$) belong to representations of the subgroup $ U(5) \subset SO(10)$. Our  distinction  between physical  and  un--physical  sectors  refers  to the  Euclidean  framework.  In this case  the  states  involving  10  bosonic fields $ x^m$ and  the  physical fermionic fields  $ \theta_{L/R}^{ph}$ and $ d_{L/R}^{ph}$ have  positive  norm and $ {\cal H}_{ph} $ is the physical Hilbert space.   When the Wick rotation is reversed to get the Lorentz version with $SO(1,9)$ as the structure group, $U(5)$ becomes one of its non compact versions and the  physical  space further reduces to be formed of 8 bosonic and 8 fermionic fields, as  will be  discussed  in the  last section  of  the  paper.

The Lagrangian ({\ref{1}}) can be obtained as follows. One starts from the Green--Schwarz Lagrangian
 \begin{eqnarray}
 L_{GS} ={1 \over 2} (\theta_{L}\fsl{ e_{-}}\nabla_{+}\theta_{L})   + {1 \over 2} (\theta_{R}\fsl{ e_{+}} \nabla_{-}\theta_{R}) - {1 \over 2}( \theta_{L}\fsl{e_{-}}\gamma_{\ast} \fsl{e_{+}}\theta_{R}).
  \label{8}
  \end{eqnarray}
 Notice that its variation is $$ \delta L_{GS} = (\delta \theta_{L} \fsl{e_{-}}[\nabla_{+}\theta_{L} - {1 \over 2} \gamma_{\ast}\fsl{e_{+}}\theta_{R}]) +  (\delta \theta_{R} \fsl{e_{+}}[\nabla_{-}\theta_{R} - {1 \over 2} \gamma_{\ast}\fsl{e_{-}}\theta_{L}])$$ so  that $L_{GS}$ is invariant under the $\kappa$--symmetry
 $ \delta \theta_{L} = k_{L}\fsl{e_{-}}$, $ \delta \theta_{R} = k_{R}\fsl{e_{+}}$ as  a  consequence  of  the Virasoro constraints  $ (e_{-}^{a}e_{- a}) = 0  = (e_{+}^{a}e_{+ a})$.
  Then one adds  the new term
 \begin{eqnarray}
   L_{new} = - ( d_{L,-}K_{L}(\nabla_{+}\theta_{L} - {1 \over 2}\gamma_{\ast}\fsl{e_{+}}\theta_{R} )) -
   ( d_{R,+}K_{R}(\nabla_{-}\theta_{R} + {1 \over 2}\gamma_{\ast}\fsl{e_{-}}\theta_{L}))
\nonumber
\end{eqnarray}
\begin{eqnarray}
- {1 \over 2}( d_{L,-}K_{L}\gamma_{\ast}\tilde K_{R} d_{R,+}).
\label{9}
\end{eqnarray}
The  addition  of  $L_{new} $  promotes  the  $\kappa$--symmetry to a  BRS symmetry  involving  the  pure  spinors  $\lambda_{L}$, $\lambda_{R}$ and  $ L_{GS} + L_{new}$ is invariant under the BRS transformations ({\ref{4}}), (\ref{4bis}), ({\ref{5}}) and ({\ref{6}}).

The action $\int L $, with $L$ defined in ({\ref{1}}), is obtained by adding to $\int (L_{GS} + L_{new})$ a suitable BRS--exact  gauge fixing term
 \begin{eqnarray}
 \int L_{gf} = s \int F_{gf}
\label{10}
\end{eqnarray}
  where $F_{gf}$  is the so called gauge fermion of  ghost  number $n_{gh} = - 1 $.
Choosing
\begin{eqnarray}
F_{gf} = -( \omega_{L,-}\nabla_{+} \theta_{L})   - ( \omega_{R,+}\nabla_{-} \theta_{R}) +
{1 \over 2} (\omega_{L,-}\gamma_{\ast} \fsl{e_{+}}\theta_{R}) - {1 \over 2} (\omega_{R,+}\gamma_{\ast}  \fsl{e_{-}} \theta_{L})
\nonumber
\end{eqnarray}
\begin{eqnarray}
- {1 \over 4}[\omega_{L,-} \gamma_{\ast}(1 - \tilde K_{R}) d_{R,+} -
\omega_{R,+} \gamma_{\ast}(1 - \tilde K_{L}) d_{L,-}]
\label{11}
\end{eqnarray}
one  gets
\begin{eqnarray}
L_{gf} = s F_{gf} =   -( \omega_{L,-}\nabla_{+} \lambda_{L})   - ( \omega_{R,+}\nabla_{-} \lambda_{R})
\nonumber
\end{eqnarray}
\begin{eqnarray}
 -( d_{L,-}(1 -  K_{L})\nabla_{+} \theta_{L})   - ( d_{R,+}(1 - K_{R})\nabla_{-} \theta_{R})
\nonumber
\end{eqnarray}
\begin{eqnarray}
+ {1 \over 2} (d_{L,-}(1 -  K_{L})\gamma_{\ast} \fsl{e_{+}} \theta_{R}) - {1 \over 2} (d_{R,+}(1 -  K_{R})\gamma_{\ast} \fsl{e_{-}} \theta_{L})
\nonumber
\end{eqnarray}
\begin{eqnarray}
- {1 \over 2} (d_{L,-}(1 - K_{L}) \gamma_{\ast}(1 - \tilde K_{R}) d_{R,+})\,.
\label{11a}
\end{eqnarray}
With this choice any dependence on $K_{L}$ and  $K_{R}$ disappears from the total action  and $ L_{GS} + L_{new} + L_{gf} $ coincides with $ L $ in ({\ref{1}}).
Projected  with $K_{L/R}$, the  first  two terms of $L_{GS}$, neglecting a total  derivative,  give
\begin{eqnarray}
{1 \over 2} (\theta_{L/R}\fsl{e_{\mp}}\nabla_{\pm}\theta_{L/R}) = ( \theta_{L} \fsl{e_{\mp}} K_{L/R}\nabla_{\pm}(K_{L/R}\theta_{L/R}))
\nonumber
\end{eqnarray}
\begin{eqnarray}
+ {1 \over 2}(\theta_{L/R}(1 - \tilde K_{L/R})\fsl{e_{\mp}}(1 - K_{L/R})\nabla_{\pm}((1 - K_{L/R}) \theta_{L/R})) + L_{L/R}^{(1)}
\label{12}
\end{eqnarray}
and the last term  of $L_{GS}$ yields
\begin{eqnarray}
- {1 \over 2}( \theta_{L}\fsl{e_{-}}\gamma_{\ast} \fsl{e_{+}}\theta_{R}) = -{1 \over 2} ( \theta_{L}\tilde K_{L} \fsl{e_{-}}\gamma_{\ast}\fsl{e_{+}}K_{L}\theta_{R})
- {1 \over 2}( \theta_{L}\fsl{e_{-}}K_{L} \gamma_{\ast}\tilde K_{R}\fsl{e_{+}} \theta_{R})
\nonumber
\end{eqnarray}
\begin{eqnarray}
  - {1 \over 2} ( \theta_{L}(1 - \tilde K_{L}) \fsl{e_{-}}(1 -  K_{L})  \gamma_{\ast}
( 1 - \tilde K_{R})\fsl{e_{+}}(1 - K_{R}) \theta_{R}) +  L^{(2)}\,,
\label{12a}
\end{eqnarray}
where
\begin{eqnarray}
L_{L/R}^{(1)} = (\theta_{L/R}(1 - {1 \over 2} \tilde K_{L/R})\fsl{e_{\mp}}\nabla_{\pm}( K_{L/R})K_{L/R} \theta_{L/R})
\nonumber
\end{eqnarray}
\begin{eqnarray}
- {1 \over 2 }(\theta_{L/R}(1 - \tilde K_{L/R})\fsl{e_{\mp}}\nabla_{\pm}( K_{L/R})(1 - K_{L/R}) \theta_{L/R})
\label{12b}
\end{eqnarray}
and
\begin{eqnarray}
L^{(2)} =  - {1 \over 2} ( \theta_{L} \tilde K_{L} \fsl{e_{-}} \gamma_{\ast}\fsl{e_{+}}(1 - K_{R}) \theta_{R})   - {1 \over 2} ( \theta_{L}(1 - \tilde K_{L}) \fsl{e_{-}} \gamma_{\ast}
\fsl{e_{+}} K_{R} \theta_{R})
\label{13c}
\end{eqnarray}

Then adding  to $L_{GS}$ the terms $L_{new}$ given  in (\ref{9}) and $L_{gf}$ given in (\ref{11a}), and
using  the  definitions (\ref{7}), (\ref{7a}) and (\ref{7b}), as well as
\begin{eqnarray}
  \theta^{unph}_{L} = (1 -K_{L}) \theta_{L},  \qquad \theta^{unph}_{L} = (1 -K_{L}) \theta_{L}
\label{14}
\end{eqnarray}
\begin{eqnarray}
 d^{unph}_{L,-} = [d_{L,-} - {1 \over 2}\theta^{unph}_{L}\fsl{e_{-}}](1 - K_{L}),
\nonumber
\end{eqnarray}
\begin{eqnarray}
d^{unph}_{R,+} = [d_{R,+} - {1 \over 2}\theta^{unph}_{R}\fsl{e_{+}}](1 - K_{R}),
\label{14a}
\end{eqnarray}
one obtains
\begin{eqnarray}
L_{GS} + L_{new} + L_{gf} = L^{ph} + L^{unph} + L^{(1)} + L^{(2)} +
\nonumber
\end{eqnarray}
\begin{eqnarray}
 - {1 \over 2} [(d_{L,-}(1 -  K_{L})\gamma_{\ast} \fsl{e_{+}} \theta_{R}) -  (d_{R,+}(1 -  K_{R})\gamma_{\ast} \fsl{e_{-}} \theta_{L})]
\nonumber
\end{eqnarray}
\begin{eqnarray}
+ {3  \over 2} (d_{L,-}(1 - K_{L}) \gamma_{\ast}(1 - \tilde K_{R}) d_{R,+})\,,
\label{15}
\end{eqnarray}
where
 \begin{eqnarray}
  L^{ph}= - [ (d_{L,-}^{ph}\nabla_{+}\theta^{ph}_{L}) + (d_{R,+}^{ph}\nabla_{-}\theta^{ph}_{R}) +
  {1 \over 2}( d_{L,-}^{ph} \gamma_{\ast} d_{R,+}^{ph}) +
 \nonumber
 \end{eqnarray}
 \begin{eqnarray}
 {1 \over 2} ( \theta^{ph}_{L} \fsl{e_{-}}\gamma_{\ast}\fsl{e_{+}}
\theta^{ph}_{R}) ]
\label{16}
\end{eqnarray}
\begin{eqnarray}
L^{unph}  = -( d^{unph}_{L,-}\nabla_{+} \theta^{unph}_{L})   - (d^{unph}_{R,+}\nabla_{-} \theta^{unph}_{R})
\nonumber
\end{eqnarray}
 \begin{eqnarray}
- 2 ( d^{unph}_{L,-}\gamma_{\ast}d^{unph}_{R,+})  -( \omega_{L,-}\nabla_{+} \lambda_{L})   - ( \omega_{R,+}\nabla_{-} \lambda_{R})
\label{17}
\end{eqnarray}
and
\begin{eqnarray}
L^{(1)} = L_{L}^{(1)} +  L_{R}^{(1)}
\label{14b}
\end{eqnarray}
where $ L_{L/R}^{(1)}$ are  defined  in (\ref{12b}).

For  the  purpose of  this  paper it  is  not  relevant to have  an action which would be independent of  $K_{L/R}$ and therefore  there is  a large  freedom  in  the choice of  $L_{gf}$. In  other  words,  adding BRS  exact  terms to  $L$  gives rise to  Lagrangians  $L'$ which are equivalent  to  $L$.
Of course  the  terms  in  the last two   rows  of (\ref{11a}) (as  well as  the last two  terms  in (\ref{15})) are, by construction, BRS exact and  can  be  subtracted. (Those  of  the  first  two rows of (\ref{11a})  must  be retained  if one  insists to  have  propagating  ghosts  $\lambda_{L/R}$ and $\omega_{L/R}$).

 Moreover a  BRS  invariant  term which  contains   $ \lambda_{L} \fsl{e_{-}}$ or $ \lambda_{R} \fsl{e_{+}}$ is  BRS  exact,  being  the  BRS  variation  of the  same  term  with $ \lambda_{L/R} \fsl{e_{\mp}}$  replaced by  $d_{L/R}K_{L/R}$. For instance, the  first term $ L^{(2)}$ in  (\ref{13c}) is  BRS exact due  to  the  identity
\begin{eqnarray}
 ( \theta_{L} \tilde K_{L} \fsl{e_{-}} \gamma_{\ast}\fsl{e_{+}}(1 - K_{R}) \theta_{R})  = \frac{(\theta_{L}\gamma_{\ast}
\Gamma_{a}\gamma_{\ast}\lambda_{L})(\lambda_{R}\fsl{e_{+ }} \Gamma^{a}\fsl{v_{-}} (1 - K_{R})\theta_{R})}{ {2(\lambda_{L}\gamma_{\ast} \lambda_{R})}}
\label{17a}
\end{eqnarray}
where $ \fsl{v}= v^{a}\Gamma_{a} $ and $v^a $ is the one--form with components
\begin{eqnarray}
v_{\pm}^{a} \equiv ( e_{\pm}^{\tilde a}, - e_{\pm}^{i})
\label{17b}
\end{eqnarray}
so that
$$  \gamma_{\ast} \fsl{e_{\pm}} = \fsl{v_{\pm}} \gamma_{\ast}.$$
A similar  identity holds for  the  second  term of (\ref{13c}), and  therefore $L^{(2)}$  can  be subtracted  from  the  action.
The  same happens  for  BRS  invariant  terms that  contain  a factor $ \nabla_{\pm}\lambda_{L}$ or $ \nabla_{\pm}\lambda_{R}$ which are  the BRS variation  of  the  same terms  with $\nabla_{\pm}\lambda_{L}$ replaced  by $ \nabla_{\pm}(1 - K_{L/R}) \theta_{L/R}$. In  particular,  terms like $L_{L/R}^{(1)}$ in (\ref{12b}) ( or $L^{(1)} $ in (\ref{14b})) that  contain  the factors $\nabla_{\pm}K_{L/R}$  are  BRS exact.

However  this  claim requires a clarification. Indeed  it  stays on the implicit assumption that $\nabla_{\pm}K_{L/R}$  produces only  terms  proportional  to $\nabla_{\pm}\lambda_{L/R}$ i.e. that
$ s \nabla_{\pm} \gamma_{\ast} = 0 $. Since $\gamma_{\ast}$  is the related to the value of the Ramond--Ramond flux in $AdS_{5}\times S_{5}$ which  connects  left--handed and  right--handed  spinors, and the pure spinor formulation allows for different left--handed and right--handed Lorentz (and Weyl) connections, one might  assume  that  the  BRS variation of  $  \nabla \gamma_{\ast}  $ vanishes only if  the left--handed and right--handed Lorentz connections are equal.  However   $ L^{(1)}$  can  be  BRS  trivial  also  if they  are  different but satisfy a certain condition. If the connections are different
\begin{eqnarray}
\nabla_{\pm}\gamma_{\ast} = \fsl{\Omega}_{L,\pm } \gamma_{\ast} + \gamma_{\ast}
 \fsl{\Omega}_{R,\pm}= ( \fsl{\Omega}_{L,\pm} - \fsl{\tilde \Omega}_{R,\pm})\gamma_{\ast} =  \fsl{\Omega}_{\pm}\gamma_{\ast}
 \label{17bis}
 \end{eqnarray}
 where $ \tilde \Omega_{L/R} =  \gamma_{\ast}\Omega_{L/R}\gamma_{\ast}$,  $  \fsl{\Omega}_{L/R}= {1 \over 4} \Omega_{L/R}^{[ab]}\Gamma_{ab}$   are (classical) left-handed and right-handed Lorentz connections and  $ \fsl{\Omega}= ( \fsl{\Omega}_{L} - \fsl{\tilde \Omega}_{R})$. If $ \fsl{\Omega}\ne 0$,
 the  terms in the  second line  of $L^{(1)}_{L/R}$ in (\ref{12b}) are still  BRS exact and the first  terms are exact
 provided that
 \begin{eqnarray}
  \fsl{e_{\mp}}\fsl{\Omega}_{\pm} = 0.
  \label{17tris}
  \end{eqnarray}
 As we will see, this condition  is satisfied by the connection $\fsl{\Omega}$ that  we  will need  for consistency.

Therefore,  neglecting  BRS  exact terms, the  action $\int L $  is  equivalent  to  the action
\begin{eqnarray}
\int L' = \int L^{ph} + \int L^{unph}
\label{18}
\end{eqnarray}

$L^{ph}$ describes ( in  the  Euclidean signature)  the  physical, fermionic,  sector  and  $ L^{unph}$  desribes  the  unphysical  sector.

The Lagrangian of  the physical  sector describes a set of 5 + 5 left-handed and 5 + 5 right-handed physical fields with field equations that are linear in derivatives. This is in agreement with the number of the  10 physical bosonic fields $x^{m}$ (5 for the Euclidean version of $ AdS_{5}$ and 5 for $ S_{5}$) with the field equations which are quadratic in derivatives.

As it  will be seen in the  next  section, in the case of the Lorentzian  signature
 this result also agrees with that of the Green-Schwarz formulation, where  there are  8 + 8 fermionic  physical  fields  and 2 + 2 Grassmann--odd degrees of freedom  provided by the b--c ghosts.

As  for  the  unphysical  sector  described by the action $\int L^{unph}$, it follows from
 eqs. ({\ref{4}}) and ({\ref{5}}) that $\omega_{L,-}$, $\omega_{R,+}$ and $\theta^{unph}_{L}$, $\theta^{unph}_{R}$ do not belong to the physical Fock space since they are not BRS invariant and $\lambda_{L}$, $\lambda_{R}$  and $ d_{L,-}(1-K_{L})$, $d_{R,+}(1-K_{R})$ , being BRS exact, are swept away by the quotient that defines the physical space ${\cal H}_{ph}$. Moreover  it  follows  from  (\ref{17}) by  computing  the functional  determinant that $ d^{unph}_{L/R}$, $\theta^{unph}_{L/R}$, $\lambda_{L/R}$ and $\omega_{L/R}$ are  massless
  and  form 11 left-handed and  11 right-handed  BRS  quartets.
Therefore  they  give  a vanishing  contribution  to  the  one  loop  partition  function, since the  contributions  of  the  members  of  each  quartet cancel  each other.

\section{\bf The  Physical Fermionic Sector of the PS Action in $AdS(5)\times S(5)$ }

Now  let  us  discuss the action $\int L^{ph}$  for  the  physical  sector, in order  to  clarify  the  relation of this free action and its spectrum  to  the corresponding action  and  spectrum  of  the
Green-Schwarz  theory.  To this end  we shall  work  in the Lorentzian  frame.

The  action  $ \int  L^{ph} $ in  ({\ref{16}) is  similar  to  the gauge  fixed  Green-Schwarz action  ({\ref{011}) but  with  a  relevant  difference. The  GS  action describes 8 left-handed  and  8 right-handed fermionic  fields whereas the  physical  action $\int L^{ph}$  describes 5+5 left-handed   and  5+5  right-handed fermions.  Therefore, in  order to  compare them  one  needs projectors $\Pi^{(4)}_{L/R}$ and $\Pi^{(1)}_{L/R}$ that  commute  with $K_{L/R}$ and among  themselves  and  which project  on  a  4-dimensional  and a 1-dimensional subspace respectively.
The projectors satisfy  the  conditions
\begin{eqnarray}
\Pi^{(4)}_{L/R} + \Pi^{(1)}_{L/R} = K_{L/R},
\label{40a}
\end{eqnarray}
\begin{eqnarray}
 \Pi^{(1)}_{L/R} \Pi^{(1)}_{L/R} = \Pi^{(1)}_{L/R}, \qquad \Pi^{(4)}_{L/R} \Pi^{(4)}_{L/R}= \Pi^{(4)}_{L/R},
 \label{40b}
 \end{eqnarray}
 \begin{eqnarray}
 \Pi^{(4)}_{L/R}\Pi^{(1)}_{L/R} = 0 = \Pi^{(1)}_{L/R} \Pi^{(4)}_{L/R}.
 \label{40d}
 \end{eqnarray}
 Notice  however that  it is sufficient  that  these  properties  hold  at the cohomological level i. e.
 modulo  BRS  trivial  terms.  In  fact, since the physical  Lagrangian $L^{ph}$ depends  only  on BRS  invariant  fields (i.e. $\theta_{L/R}^{ph}$ and
 $d_{L/R}^{ph}$), then when  $\Pi^{(4)}_{L/R}$  and $\Pi^{(1)}_{L/R}$ are  used  in  $L^{ph}$, these  BRS  trivial  contributions give  rise  to  terms  in  the  action  that  are  BRS  exact  and  therefore  can  be  subtracted,  as  discussed before. Therefore, in  the following we  shall imply  that the identities (including (\ref{40b}) and (\ref{40d})) hold  modulo  BRS  trivial  terms.
We  shall  use  the  notation  $\doteq $  to  mean ``equal  modulo  BRS  exact  terms" (then  also  in (\ref{40b}) and (\ref{40d})  the  signs  of the equality should  be  replaced  with  $\doteq $).

 Let us first discuss the case when the string motion is restricted to $AdS(5)$. In this case it
is  not  difficult  to  guess the  form  of  these  projectors:
 \begin{eqnarray}
\Pi^{(4)}_{L/R} = {1 \over {2(e_{+}e_{-})}} K_{L/R}\fsl{e_{\pm}}\fsl{e _{\mp}}K_{L/R} \doteq
{1 \over {2(e_{+}e_{-})}} K_{L/R}\fsl{e_{\pm}}\fsl{e _{\mp}},
 \label{41}
 \end{eqnarray}
\begin{eqnarray}
 \Pi^{(1)}_{L/R} ={1 \over {2(e_{+}e_{-})}} K_{L/R}\fsl{e_{\mp}}\fsl{e _{\pm}}K_{L/R}\doteq K_{L/R}{
 {(\fsl{e_{\mp}}\gamma_{\ast}\lambda_{R/L})(\lambda_{L/R} \fsl{e_{\pm}})} \over {2(e_{+}e_{-})(\lambda_{L}\gamma_{\ast}\lambda_{R})}}K_{L/R},
 \nonumber
 \end{eqnarray}
 \begin{eqnarray}
  \doteq {{(\fsl{e_{\mp}}\gamma_{\ast}\lambda_{R/L})(\lambda_{L/R} \fsl{e_{\pm}})} \over {2(e_{+}e_{-})(\lambda_{L}\gamma_{\ast}\lambda_{R})}},
\label{41bis}
\end{eqnarray}
which  indeed  satisfy,  at  the  cohomological  level, the  conditions   ({\ref{40a}})-- (\ref{40d}).
Moreover $$ Tr(\pi^{(4)}_{L/R}) \doteq 4 \qquad Tr(\pi^{(1)}_{L/R})\doteq  1 $$  so  that,  at  the  cohomological level,  they  project on to a  4-dimensional and 1-dimensional  subspaces.

Now  let  us  come  back  to  the  physical  action (\ref{16}) that  we  write  in  the  form
\begin{eqnarray}
\int L^{ph} = \int [  d_{L,-}^{ph} K_{L}\nabla_{+}(K_{L}\theta_{L}) + d_{R,+}^{ph} K_{R} \nabla_{-}(K_{R}\theta_{R})
\nonumber
\end{eqnarray}
\begin{eqnarray}
- {1 \over 2} (  d_{L,-}^{ph}K_{L}\gamma_{\ast} \tilde K_{R}  d_{R,-}^{ph}) - {1 \over 2} (\theta_{L}
\tilde K_{L} \fsl{e_{-}}\gamma_{\ast} \fsl{e_{+}} K_{R} \theta_{R})]
\label{42}
\end{eqnarray}
and  use  the  identity (\ref{40a}) to  replace  $  K_{L/R} $ with
 the  projectors $ \Pi^{(4)}_{L/R}$ and $\Pi^{(1)}_{L/R}$  defined  in (\ref{41}) and (\ref{41bis}).
 Then one  gets
 \begin{eqnarray}
  d_{L,-}^{ph}\Pi^{(4)}_{L} = {1 \over {2(e_{+}e_{-})}} ( d_{L,-}^{ph}\fsl{e_{+}}) \fsl{e_{-}} K_{L} = \bar \theta_{L}^{(p)}\fsl{e_{-}} K_{L}
 \nonumber
 \end{eqnarray}
 \begin{eqnarray}
 d_{L,-}^{ph}\Pi^{(1)}_{L} \doteq {1 \over {2(e_{+}e_{-})(\lambda_{L}\gamma_{\ast}\lambda_{R})}} ( d_{L,-}^{ph} \fsl{e_{-}}\gamma_{\ast}\lambda_{R})
\lambda_{L}\fsl{e_{+}} = \hat b_{L,--}{1 \over {(e_{+}e_{-})}}\lambda_{L}\fsl{e_{+}}
\label{43a}
\end{eqnarray}
\begin{eqnarray}
 d_{R,+}^{ph}\Pi^{(4)}_{R} = {1 \over {2(e_{+}e_{-})}} ( d_{R,+}^{ph}\fsl{e_{-}}) \fsl{e_{+}} K_{R} = \bar \theta_{R}^{(p)} \fsl{e_{+}} K_{R}
\nonumber
\end{eqnarray}
\begin{eqnarray}
 d_{R,+}^{ph}\Pi^{(1)}_{R} \doteq {1 \over {2(e_{+}e_{-})(\lambda_{L}\gamma_{\ast}\lambda_{R})}} ( d_{R,+}^{ph} \fsl{e_{+}}\gamma_{\ast}\lambda_{L})\lambda_{R}\fsl{e_{-}} = \hat b_{R,++}
{1 \over {(e_{+}e_{-})}}\lambda_{R} \fsl{e_{-}}
\label{43b}
\end{eqnarray}
where  we  have  defined
\begin{eqnarray}
\bar \theta_{L}^{(p)} =  {1 \over {2(e_{+},e_{-})}} d_{L,-}^{ph}\fsl{e_{+}}, \qquad  \bar \theta_{R}^{(p)} =  {1 \over { 2(e_{+},e_{-})}} d_{R,+}^{ph}\fsl{e_{-}}
\label{44a}
\end{eqnarray}
and
\begin{eqnarray}
\hat b_{L, --} = {1 \over {2(\lambda_{L}\gamma_{\ast} \lambda_{R})}}
 (\hat d_{L,-}\fsl{e_{-}}\gamma_{\ast}\lambda_{R}), \qquad
\hat b_{R,++} = {1 \over {2(\lambda_{L}\gamma_{\ast} \lambda_{R})}} (\hat d_{R,+} \fsl{e_{+}}\gamma_{\ast}\lambda_{L})
\label{44b}
\end{eqnarray}
 The  definitions (\ref{44a})  look  natural  since $\bar \theta^{(p)}_{L}$,  $\bar \theta^{(p)}_{R}$ (like $\theta_{L}$,$\theta_{R}$) have  conformal weights (0,0).
Moreover  we  define
\begin{eqnarray}
\Pi^{(4)}_{L}\theta_{L} =  \theta_{L}^{(p)}, \qquad   \Pi^{(4)}_{R} \theta_{R} = \theta_{R}^{(p)}
\label{45}
\end{eqnarray}
and
\begin{eqnarray}
\Pi^{(1)}_{L}\theta_{L} \doteq {1 \over {2(e_{-}e_{+})}}{{ (\fsl{e_{-}}\gamma_{\ast}\lambda_{R})} \over
{(\lambda_{L}\gamma_{\ast}\lambda_{R})}}(\lambda_{L}\fsl{e_{+}}\theta_{L})
= {{(\fsl{e_{-}}\gamma_{\ast}\lambda_{R})} \over {2(\lambda_{L}\gamma_{\ast}\lambda_{R})}} \hat c^{-}_{L}
\label{46a}
\end{eqnarray}
\begin{eqnarray}
 \Pi^{(1)}_{R}\theta_{R} \doteq {1 \over {2(e_{-}e_{+})}} {{(\fsl{e_{+}}\gamma_{\ast}\lambda_{L})(\lambda_{R}\fsl{e_{-}}\theta_{R})} \over {(\lambda_{L}\gamma_{\ast} \lambda_{R})}}
= {{(\fsl{e_{+}}\gamma_{\ast}\lambda_{L})}
\over { 2 (\lambda_{L}\gamma_{\ast}\lambda_{R})}} \hat c^{+}_{R}
\label{46b}
\end{eqnarray}
where
\begin{eqnarray}
\hat c^{-}_{L} = {1 \over {(e_{-}e_{+})}} (\lambda_{L}\fsl{e_{+}}\theta_{L}),  \qquad \hat c^{+}_{R} = {1 \over {(e_{-}e_{+})}}(\lambda_{R}\fsl{e_{-}}\theta_{R}).
\label{47}
\end{eqnarray}
Using  the  identity  (\ref{40a}) the  action (\ref{42}) splits  in four  parts: $\int L^{ph}_{4,4}$ when
 both $K_{L/R}$ are replaced by  $\Pi^{(4)}_{L/R}$,  $\int L_{1,1}$ when
  both $K_{L/R}$ are replaced by  $\Pi^{(1)}_{L/R}$, $\int L_{4,1}$ when
the first $K_{L/R}$ is replaced by  $\Pi^{(4)}_{L/R}$  and  the  second one by $\Pi^{(1)}_{L/R}$ and
$\int L_{1.4}$ when
the first $K_{L/R}$ is replaced by  $\Pi^{(1)}_{L/R}$  and  the  second one by $\Pi^{(4)}_{L/R}$.
 Then, modulo  BRS  exact  terms,  one  obtains
\begin{eqnarray}
L^{ph}_{4,4} \doteq (\bar \theta_{L}^{(p)}\fsl{e_{-}}\nabla_{+}\theta_{L}^{(p)}) +
 (\bar \theta_{R}^{(p)}\fsl{e_{+}}\nabla_{-}\theta_{R}^{(p)})
 \nonumber
 \end{eqnarray}
 \begin{eqnarray}
  - {1 \over 2} (\bar \theta_{L}^{(p)}\fsl{e_{-}}\gamma_{\ast} \fsl{e_{+}} \bar \theta_{R}^{(p)})
 - {1 \over 2} (\theta_{L}^{(p)} \fsl{e_{-}}\gamma_{\ast} \fsl{e_{+}} \theta_{R}^{(p)})
\label{48}
\end{eqnarray}
\begin{eqnarray}
L_{1,1} \doteq [(\hat b_{L,--}\nabla_{+}\hat c^{-}_{L}) + (\hat b_{R,++}\nabla_{-}\hat c^{+}_{R})] - (\hat b_{L,--}{\cal G}\hat b_{R,++})
\label{48a}
\end{eqnarray}
where
\begin{eqnarray}
 {\cal G} ={{(\lambda_{L} \fsl{e_{+}}\gamma_{\ast}\fsl{e_{-}} \lambda_{R})} \over {2(e_{-}e_{+})^{2}}}
\label{48bis}
\end{eqnarray}
and
\begin{eqnarray}
L_{4,1} + L_{1,4} \doteq L^{a} + L^{b} + L^{c}
\label{49}
\end{eqnarray}
where
\begin{eqnarray}
 L^{(a)} =  {1 \over {2(e_{-}e_{+})}}[ \hat b_{L,--}(\lambda_{L}\fsl{e_{+}}\gamma_{\ast} \fsl{e _{+}}
\theta_{R}^{(p)}) + \hat b_{R,++}(\lambda_{R}\fsl{e_{-}}\gamma_{\ast} \fsl{e_{-}}\theta_{L}^{(p)})]
\label{49a}
\end{eqnarray}
\begin{eqnarray}
 L^{(b)} ={1 \over {(e_{-}e_{+})}} [ \hat b_{L,--}(\lambda_{L}\fsl{e_{+}}(\nabla_{+}{1 \over {2(e_{-}e_{+})}}\fsl{e_{+}})\fsl{e_{-}}\theta_{R}^{(p)}) +
 \nonumber
 \end{eqnarray}
 \begin{eqnarray}
\hat b_{R,++}(\lambda_{R}\fsl{e_{-}}(\nabla_{-}{1 \over {2(e_{-}e_{+})}}\fsl{e_{-}})\fsl{e_{+}}\theta_{L}^{(p)}) ]
\label{49b}
\end{eqnarray}
\begin{eqnarray}
L^{(c)}= {1 \over {(e_{-}e_{+})}} [ \hat b_{L,--}(\lambda_{L}\fsl{e_{+}}(\nabla_{+}K_{L})
\theta_{R}^{(p)}) +
\hat b_{R,++}(\lambda_{R}\fsl{e_{-}}(\nabla_{-}K_{R})\theta_{L}^{(p)}) ]
\label{49c}
\end{eqnarray}

If the  string  moves only  in $ AdS_{5}$,  $\fsl{e_{\pm}}$  commute  with $\gamma_{\ast}$ so that ${\cal G}$  reduces to
\begin{eqnarray}
{\cal G } = {{(\lambda_{L}\gamma_{\ast}\lambda_{R})} \over {(e_{-}e_{+})}}
\label{49d}
\end{eqnarray}
  and $L^{(a)}$  in (\ref{49a}) vanishes.
As for $ L^{(b)}$ and $L^{(c)}$, eq. (\ref{49b}) can be rewritten as
\begin{eqnarray}
 L^{(b)} ={1 \over {(e_{-}e_{+})}} [ \hat b_{L,--}(\lambda_{L}\fsl{e_{+}}(\nabla_{+}[{{e_{+}^{[a}e_{-}^{b]}}
 \over {(e_{-}e_{+})}}]{1 \over 2}\Gamma_{ab} )\theta_{R}^{(p)})
 \nonumber
 \end{eqnarray}
 \begin{eqnarray}
+ \hat b_{R,++}(\lambda_{R}\fsl{e_{-}}(\nabla_{-}[{{e_{-}^{[a}e_{+}^{b]}}\over {(e_{-}e_{+})}}]{1 \over 2} \Gamma_{ab})\theta_{L}^{(p)}) ]
\label{49e}
\end{eqnarray}
 and, according  to  the  discussion before eq. (\ref{17bis}), $L^{(c)}$  can  be  rewritten as
\begin{eqnarray}
L^{(c)}= {1 \over {(e_{-}e_{+})}} [ \hat b_ {L,--}(\lambda_{L}\fsl{e_{+}}\fsl{\Omega_{+}}
\theta_{R}^{(p)}) +
\hat b_ {R,++}(\lambda_{R}\fsl{e_{-}}\fsl{\Omega_{-}}\theta_{L}^{(p)}) ]
\label{50}
\end{eqnarray}
where  the  Lorentz  valued one form  $\fsl{\Omega}$ is the difference between the left-handed and right-handed  Lorentz  connections.
Then, if for consistency we  take for $ \fsl{\Omega_{\pm}}$
 \begin{eqnarray}
 \fsl{\Omega_{\pm}} = - \nabla_{\pm}[{{e_{\pm}^{[a}e_{\mp}^{b]}}\over {(e_{-}e_{+})}}]{1 \over 2} \Gamma_{ab}
\label{56}
\end{eqnarray}
  $L^{(c)} $ and $L^{(b)}$ in (\ref{49e}) and (\ref{50}) cancel
and  $ L^{ph}_{4,1} + L^{ph}_{1,4}$  vanishes  at cohomological  level.
Therefore the action $\int L^{ph} $ is equivalent to the action
$$ \int L' = \int L^{ph}_{4,4}  +\int  L_{1,1} $$
where $ L^{ph}_{4,4}$ and  $ L_{1,1}$ are defined in  (\ref{48}) and (\ref{48a}).

Notice  that $ \fsl{\Omega_{\pm}}$ defined in (\ref{56}) indeed satisfies  the  condition (\ref{17tris}), as
anticipated.

As  the  Green-Schwarz action, the action $\int L^{ph}_{4,4}$ describes a set  of 8 left-handed and 8 right-handed ``massive" fermions with conformal weight (0,0). This action is similar  but  not identical to the Green-Schwarz action. However  these  two actions  have the same spectrum, as can be seen by showing that their  functional  determinants  are equal. As  for  $ \int  L_{1,1} $, the functional  determinant of  this  action  does  not  involve ${\cal  G}$ being  proportional to $ \nabla_{+}\nabla_{-}\nabla_{+}\nabla_{-} $  and therefore   $ \int  L_{1,1} $  describes  a massless b-c system.

It  is  not  difficult to  slightly modify   the  procedure to cover the  general  case in which the  string moves  in the whole $AdS(5)\times S(5)$
 \footnote { Our procedure holds for a generic motion of the string with the exception  of those singular motions where $ (v_{\pm}e_{\mp}) = 0$. }.
 For that purpose  let  us consider  the one--form $v^{a}$ defined in (\ref{17b}).
Let  us recall that
\begin{eqnarray}
 \gamma_{\ast} \fsl{e_{\pm}} = \fsl{v_{\pm}} \gamma_{\ast}
 \label{58}
 \end{eqnarray}
 and also notice that
 \begin{eqnarray}
 (v_{\pm}e_{\mp}) = (e_{\pm}v_{\mp})  \qquad  (v_{+}v_{+}) = 0 = (v_{-}v_{-})
 \label{58bis}
 \end{eqnarray}
  Then define  the  modified  projectors
 \begin{eqnarray}
\hat \Pi^{(4)}_{L/R} = {1 \over {2(v_{\pm}e_{\mp})}}K_{L/R}\fsl{v_{\pm}}\fsl{e_{\mp}}K_{L/R},
 \nonumber
 \end{eqnarray}
\begin{eqnarray}
 \hat \Pi^{(1)}_{L/R} = {1 \over {2(v_{\pm}e_{\mp})}}  (\fsl{e_{\mp}}\gamma_{\ast}\lambda_{R/L})(\lambda_{L/R}\fsl{v_{\pm}})
\label{51}
\end{eqnarray}
They satisfy, at  the  cohomological level, the  same  conditions (\ref{40a})-(\ref{40d}) satisfied  by  $ \Pi^{(4)}_{L/R}$ and $ \Pi^{(1)}_{L/R}$.
Projected  with  these modified projectors, the  analogue of (\ref{43a}), (\ref{43b}), (\ref{46a}), (\ref{46b}) are  obtained by replacing in (\ref{43a}),(\ref{46a}) $ e_{+}^{a}$ with $v_{+}^{a}$ and
 in (\ref{43b}),(\ref{46b}) $ e_{-}^{a}$ with $v_{-}^{a}$.
 The  left-handed  fields $ \bar \theta_{L}^{(p)}$, $\theta_{L}^{(p)}$, $ \hat b_{L,--}$, $\hat c^{-}_{L}$ and   the  right-handed  fields $ \bar \theta_{R}^{(p)}$, $\theta_{R}^{(p)}$, $ \hat b_{R,++}$, $\hat c^{+}_{R}$  are  changed accordingly.

For these modified fields we shall maintain the same notations as for the unmodified ones.
 In particular, $ \hat b_{L,--}$ and $ \hat b_{R,++}$ in (\ref{44b}) remain unchanged while
 $ \bar \theta_{L/R}^{(p)}$ in(\ref{44a}) become
 \begin{eqnarray}
\bar \theta_{L}^{(p)} =  {1 \over {2(v_{+},e_{-})}} d_{L,-}^{ph}\fsl{v_{+}} , \qquad  \bar \theta_{R}^{(p)} =  {1 \over { 2(e_{+},v_{-})}} d_{R,+}^{ph}\fsl{v_{-}}
\label{51bis}
\end{eqnarray}
and $\hat c^{\mp}_{L/R}$ in (\ref{47}) become
\begin{eqnarray}
\hat c^{-}_{L} = {1 \over {2(e_{-}v_{+})}} (\lambda_{L}\fsl{v_{+}}\theta_{L}),  \qquad \hat c^{+}_{R} = {1 \over {2(v_{-}e_{+})}}(\lambda_{R}\fsl{v_{-}}\theta_{R})\,.
\label{51ter}
\end{eqnarray}
As  for  the  action $\int L'$, $L^{ph}_{4,4}$ and  $L_{1,1}$ remain unchanged
but now $ {\cal G}$ becomes ${\cal G} = {{(\lambda_{L}\fsl{v_{+}}\gamma_{\ast}\fsl{v_{-}} \lambda_{R})} \over {2(e_{-}v_{+})^{2}}}$ or, using (\ref{58})  and
  modulo a BRS trivial  term,
  $$ {\cal G } = {{(\lambda_{L}\gamma_{\ast}\lambda_{R})} \over {(e_{-}v_{+})}}. $$
  which  agrees with (\ref{49d}) if the  string moves  only in  $AdS(5)$.

Now  for  $ L_{4,1} + L_{1,4}\doteq  L^{(a)} +  L^{(b)} +  L^{(c)}$ one has
\begin{eqnarray}
 L^{(a)} =   {1 \over {2(e_{-} v_{+})}}[ \hat b_{L,--}(\lambda_{L}\fsl{v_{+}}\gamma_{\ast} \fsl{e _{+}}
\theta_{R}^{(p)}) + \hat b_{R,++}(\lambda_{R}\fsl{v_{-}}\gamma_{\ast} \fsl{e_{-}}\theta_{L}^{(p)})],
\label{52}
\end{eqnarray}
\begin{eqnarray}
 L^{(b)} ={1 \over {2(e_{-}v_{+})}} [ \hat b_{L,--}(\lambda_{L}\fsl{v_{+}} [(\nabla_{+}{{v_{+}^{a}} \over {(v_{+}e_{-})}})e_{-}^{b}\Gamma_{ab}]\theta_{R}^{(p)})
 \nonumber
 \end{eqnarray}
 \begin{eqnarray}
+ \hat b_{R,++}(\lambda_{R}\fsl{v_{-}}[(\nabla_{-}{{v_{-}^{a}} \over {(v_{-}e_{+})}})e_{+}^{b}\Gamma_{ab}]\theta_{L}^{(p)}) ],
\label{53}
\end{eqnarray}
\begin{eqnarray}
 L^{(c)} =  {1 \over {(v_{+}e_{-})}} [ \hat b_{L,--}(\lambda_{L}\fsl{v_{+}}\fsl{\hat \Omega_{+}}
\theta_{R}^{(p)}) +
\hat b_{R,++}(\lambda_{R}\fsl{v_{-}}\fsl{\hat \Omega_{-}}\theta_{L}^{(p)}) ].
\label{54}
\end{eqnarray}
As before $L^{(a)}$ vanishes as a consequence of (\ref{58}) and (\ref{58bis}),  and $ L^{(b)}$ and $L^{(c)}$ cancel each other by choosing 
\begin{eqnarray}
\fsl{\hat \Omega_{\pm}} = - \nabla_{\pm}[{{v_{\pm}^{[a}e_{\mp}^{b]}}\over {(v_{-}e_{+})}}]{1 \over 2} \Gamma_{ab}
\label{561}
 \end{eqnarray}
which  also  satisfies  the  condition (\ref{17tris}).
Therefore, as before, $ L_{4,1} + L_{1,4}$  vanishes  at the cohomological  level.

Then the action $\int L $ is equivalent to the action
$$ \int L' = \int L^{ph}_{4,4}  + \int L_{1,1} $$
where $ L^{ph}_{4,4}$ and  $ L_{1,1}$ are defined in  (\ref{48}) and (\ref{48a}).

The action $\int L^{ph}_{4,4}$ describes a set  of 8 left-handed and 8 right-handed, massive, fermions with conformal weight (0,0) and the same ``mass" of the fermions in the GS action. Indeed (\ref{48}) reduces to  eq. (\ref{011}) with the redefinitions $$ \theta^{(+)}_{L} ={e^{{- i \pi} \over 4}}
 {{\theta^{(1)}_{L}} \choose {\theta^{(2)}_{L}}} \, , \qquad \theta^{(-)}_{R} = {e^{{i \pi} \over 4}}
 {{\theta^{(1)}_{R}} \choose {\theta^{(2)}_{R}}}  $$ where
 $$ \theta^{(1)}_{L/R} = {e^{{\pm \pi} \over 4}}(\theta^{(p)}_{L/R} - \bar \theta^{(p)}_{L/R}) \,,\qquad
  \theta^{(2)}_{L/R} = {e^{{\mp \pi} \over 4}}(\theta^{(p)}_{L/R} + \bar \theta^{(p)}_{L/R}).$$
As  for the  action $\int L_{1,1}$, it describes  a left-handed and a right-handed pair of massless anticommuting scalars with  conformal  weights (2,0), (-1,0), (0,2) and (0,-1) which are equivalent  to the b-c ghost system  of the Green-Schwarz  approach.

As the last  remark,  it  is  interesting  to notice  the  relation  between  our fields $ \hat b_{L/R,\mp \mp}$
and  the b-fields  \cite{berkquater,berkmaz,maz} for the pure spinor string theory in  $AdS(5) \times S(5)$.
Indeed, in our notation the $AdS(5) \times S(5)$ b--fields  are
\begin{equation}\label{b}
 b = {1 \over {(\lambda_{L}\gamma_{\ast} \lambda_{R})}}[{1 \over 2}
 (E_{R,-}\gamma_{\ast}E_{-}^{a}\Gamma_{a}\gamma_{\ast}\lambda_{R}) + {1 \over 4 } (E_{R,-}N_{L,-}^{ab}\Gamma_{ab}\lambda_{R}) +  {1 \over 4}(E_{R,-}j_{L,-}\lambda_{R})]
 \end{equation}
  and
  \begin{equation}\label{barb}
\bar b = {1 \over {(\lambda_{L}\gamma_{\ast} \lambda_{R})}}[{1 \over 2} ( E_{L,+} E_{+}^{a}\Gamma_{a}\gamma_{\ast}\lambda_{L}) + {1 \over 4 } (E_{L,+}N_{R,+}^{ab}\Gamma_{ab}\lambda_{L}) +
 {1 \over 4}(E_{L,+}j_{R,+}\lambda_{L})]\,
  \end{equation}
  where $ N_{L/R,\mp} = ( \omega_{L/R\mp}\Gamma^{ab}\lambda_{L/R})$ and $ j_{L/R,\mp} = ( \omega_{L/R\mp}\lambda_{L/R})$ and one recovers our expressions for $ b_{L/R, \mp \mp}$
 by neglecting the last  two  terms in \eqref{b} and \eqref{barb} (which are of higher order), replacing  $ E_{\pm}^{a}$ with the classical solutions $ e_{\pm}^{a}$ and using the field equations of $d_{L/R}$.
 Also let us notice  that  in eq. (3,35) of \cite{berkquater} the expression  for  the  zero-modes $c_{0}$
 and $\bar c_{0}$ of the c-fields in a R-R plane-wave background seems to be related to our fields $\hat c_{L}^{-}$, $ \hat c_{R}^{+}$ in (\ref{47}) (or (\ref{51ter})) \footnote { I am grateful to Luca Mazzuccato for  this remark.}.

\vspace{5mm}
\section{\bf Conclusion}
\vspace{2mm}

In  contrast to  the  case  of  string  theories  in  flat  background \cite{bm,ak}, the
equivalence  of  the Green-Schwarz and pure spinor formulations  in  curved  backgrounds, at  the  worldsheet  quantum  level,  is  still  an  open  problem.  As noted  in  the  Introduction, the  reason  is  that  string  theories  in  curved  backgrounds  are  not  free field  theories.
This problem  can  be  addressed  perturbatively  by using, for  instance,  the  background  field  method.  The  semiclassical  approximation  amounts  to  computing and comparing the  one  loop  partition  functions  of  the  quantum  fluctuations  around  the  given  background in the GS and PS formulation, as has been  done  in  \cite{abv,dima}.  Going  beyond  the  semiclassical  approximation  is  possible  in  principle  by  computing higher--loop  contributions  but  the  calculations  become  quite  hard. It  would  be interesting  to  perform  these  calculations,  even  only  at  two  loops,  since  they  involve  the  interactions among  quantum  fluctuations which  are  clearly  different  in  the two  approaches (they also affect   the  bosonic  sector).

In  this  paper, we have  proved,  at  the  level  of  the  semiclassical  approximation the equivalence  of the Green-Schwarz and pure spinor formulations of a string moving in $AdS(5) \times S(5)$ with the  method that  allows for a clear separation  of  the physical and unphysical  fermionic sectors  of  the  pure spinor  formulation.  It  has  been  show that the  unphysical  fluctuations amount  to  eleven  left--handed and  eleven  right--handed massless (bosonic and fermionic) BRS  quartets
whose  contributions  to  the  one--loop  partition  function  cancel   and that  the  quadratic physical fluctuations  in  the  two approaches have not  only  the same  spectrum  but  also  the  same conformal  weights.
This  result  has  also  revealed  an  interesting   connection  between the  fields $ \hat b_{L, --} $  and $ \hat b_{R,++}$ defined
in (\ref{44b}) and  the  b--fields  of  the  pure  spinor  approach in  $AdS(5) \times S(5)$ \cite{berkquater,berkmaz,maz}.

\begin{flushleft}
{\bf Acknowledgements.}
I am grateful to Dima Sorokin for having raised  my  interest to  this  problem and for many useful discussions and important advices  and  to  Luca Mazzuccato for a clarifying discussion and  useful  comments.
\end{flushleft}


\end{document}